\documentclass[final,5p,times,twocolumn,authoryear]{elsarticle}

\usepackage[T1]{fontenc}
\usepackage[utf8]{inputenc}
\usepackage{amsmath,amssymb}
\usepackage{booktabs}
\usepackage{array}
\usepackage{tabularx}
\usepackage{siunitx}
\usepackage{microtype}
\usepackage{url}
\usepackage{hyperref}
\usepackage{placeins}
\usepackage{float}
\usepackage{dblfloatfix} 
\usepackage{xcolor}
\usepackage{fancyhdr}
\usepackage{lastpage}
\hypersetup{hidelinks}

\graphicspath{{Figures/}}

\definecolor{astroblue}{RGB}{0,72,145}

\hypersetup{
    colorlinks=true,
    linkcolor=astroblue,
    citecolor=astroblue,
    urlcolor=astroblue
}

\newcommand{\argmax}{\operatorname*{arg\,max}}

\newcommand{\Prob}{\operatorname{P}}

\newcolumntype{L}{>{\raggedright\arraybackslash}X}
\newcolumntype{Y}{>{\centering\arraybackslash}X}

\fancypagestyle{plain}{
    \fancyhf{}

    \fancyhead[R]{%
        \footnotesize\itshape
        Extremes of solar spectral irradiance in the SORCE/XPS record
    }

    \fancyfoot[L]{%
        \footnotesize
        \textit{Preprint. July 31, 2026}
    }

    \fancyfoot[R]{%
        \footnotesize
        \thepage/\pageref*{LastPage}
    }

}

\journal{Astronomy \& Computing}

\begin{document}

\raggedbottom

\begin{frontmatter}

\title{\textbf{Extremes of solar spectral irradiance in the SORCE/XPS record}}

\author[unb]{Enzo Brasil\corref{cor1}}
\ead{enzoportobrasil@gmail.com}
\author[unb]{Cira E. G. Otiniano}
\author[unb]{Carolyne Brito}
\author[unb]{Beatriz Albernaz}
\author[ufrn]{Fidel Morales}

\cortext[cor1]{Corresponding author}

\affiliation[unb]{organization={Institute of Exact Sciences, Department of Statistics, University of Bras\'ilia, Brazil}}
\affiliation[ufrn]{organization={Center for Exact and Earth Sciences, Department of Statistics, Federal University of Rio Grande do Norte, Brazil}}


\begin{abstract}
Extreme and rare changes in space mission solar irradiance records are scientifically
relevant but difficult to quantify because these records are finite, instrument dependent, and affected by observational gaps and time varying measurement quality. We evaluated extreme daily logarithmic changes in the band integrated 0.1-7.0~nm irradiance measured by photodiode~7 of the Solar Radiation and Climate Experiment/X-Ray Photometer System (SORCE/XPS) from 2005 to 2019.
After constructing a regular daily series by linear interpolation, we analyzed
daily logarithmic changes in irradiance and fitted stationary Generalized
Extreme Value models to 60 day block maxima and transformed 15 day block
minima. Block lengths were selected using Ljung-Box diagnostics and sample
autocorrelation functions. Maximum likelihood estimation was used as the
primary inferential method, with probability weighted moments as a sensitivity
check. The maximum likelihood GEV shape estimates were $0.1661$ for maxima and $0.2928$ for transformed minima, with closely aligned
estimates under the two fitting methods. These positive point estimates are
compatible with Fr\'echet type tails under the selected block constructions. Annualized return levels provide interpretable summaries
of extreme relative increases and reductions, but estimates for long return periods remain strongly dependent on extrapolation beyond the 15 year record.
Reported measurement precision and absolute uncertainty were used to qualify
the interpretation of the fitted tails and were not propagated through the likelihood. By combining EVT based tail modeling with explicit consideration of
measurement precision, absolute uncertainty, interpolation, and mission data gaps, the analysis provides an uncertainty aware astrostatistical
baseline for extreme value inference from processed solar mission records.
\end{abstract}


\end{frontmatter}


\section{Introduction}
\label{introduction}

Solar irradiance is a central quantity in solar physics and space science, as it measures the electromagnetic energy emitted by the Sun and incident near Earth. Its temporal variability is relevant for studies of solar activity, upper-atmosphere forcing, space weather disturbances, satellite operations, and long term radiative forcing \citep{frohlich2006solar,gomez2018irradiancia,woods2021overview}.

For this reason, irradiance records must be interpreted not only as traces of
solar variability, but also as products of specific observing systems.
Instrument calibration, stability, temporal coverage, data gaps, and reported
measurement uncertainty condition the statistical conclusions that can be
drawn from these records, particularly when the analysis focuses on rare and
extreme events. \citet{eadie2019realizing} explain that this perspective is central to astrostatistics, which provides an interdisciplinary framework for adapting statistical methods to the observational, instrumental, and inferential structure of astronomical data.

The Solar Radiation and Climate Experiment (SORCE) was a NASA satellite mission developed to monitor solar irradiance, including total solar irradiance (TSI) and its distribution across different wavelengths, known as solar spectral irradiance (SSI).

It operated from 2003 to 2020, far exceeding its five-year prime mission, and contributed to the continuation of the multi-decadal TSI record, the ultraviolet SSI record, and the extension of calibrated SSI measurements across a broad spectral range \citep{rottman2005sorce,woods2021overview}. As described by \citet{woods2005xps_overview,woods2005xps_variations} and later summarized in the final SORCE/XPS data release by \citet{woods2022xpsfinal}, the X-Ray Photometer System (XPS) was designed to measure solar soft X-ray and extreme-ultraviolet (EUV) irradiance using broadband photometers. These channels sample irradiance variability associated with emission from coronal plasma and the solar transition region.

The channel analyzed here is a photodiode which monitors irradiance from 0.1 to 7.0 nm. \citet{woods2005xps_overview} describe XPS as a broadband photometric system designed to measure solar irradiance in the soft X-ray and extreme-ultraviolet ranges, while the final SORCE/XPS data release by \citet{woods2022xpsfinal} provides the processed irradiance products used in this study. The physical motivation for using this band comes from solar UV and X-ray spectroscopy and from the physics of the solar corona. \citet{phillips1945ultraviolet} describe this spectroscopy case as a central observational route for probing the solar atmosphere, including its thermal structure, dynamics, and chemical composition. \citet{aschwanden2006physics} provides the broader coronal context, in which soft X-ray and EUV observations are linked to hot coronal plasma, magnetic activity, and solar flares. In this framework, the 0.1 to 7.0 nm XPS band is physically meaningful for tail analysis because it records energetic short wavelength variability associated mainly with coronal emission and flare activity. This physical interpretation is also consistent with the review by \citet{del2018solar}, who discuss XUV and X-ray observations as diagnostics of optically thin plasma in the outer solar atmosphere.

The goal of this paper, however, is not to identify individual solar events or to infer a causal mechanism for any particular fluctuation. Instead, we address a statistical question: how can the tail behavior of a SORCE/XPS irradiance record be quantified while remaining explicit about measurement uncertainty, interpolation, and observational gaps?

Extreme solar and solar-terrestrial events have motivated a growing literature on the frequency, severity, and uncertainty of rare space-weather phenomena \citep{watari2001bastille,tsurutani2006extreme,cliver2022extreme}. Tail behavior is especially useful in this context because extreme fluctuations are poorly represented by models designed for the central part of a distribution. Extreme Value Theory (EVT) provides a natural framework for modeling extreme observations and obtaining risk measures \citep{fisher1928limiting,gnedenko1943,von1936distribution,jenkinson1955frequency,kotz2000extreme}. For an instrumental irradiance record, EVT must be interpreted jointly with data quality, as an extreme point is not only a solar observation, but also a measurement produced by a specific detector, calibration pipeline, temporal sampling scheme, and mission phase.

This paper presents a univariate EVT analysis of solar spectral irradiance from
SORCE/XPS, using the daily interpolated record from 2005 to 2019. We characterize daily logarithmic changes in irradiance, extract block maxima and transformed block minima, fit Generalized Extreme Value models, assess temporal dependence and model diagnostics, and compute annualized return levels. The contribution is both methodological and applied: we show that tail inference can complement the physical and instrumental interpretation of SORCE/XPS observations when measurement uncertainty and observational limitations are treated as features of the observational record rather than as secondary technical details.

\section{Materials and Methods}

\subsection{SORCE/XPS irradiance data}

The data analyzed in this study come from the SORCE XPS Level 3 Solar Spectral Irradiance Daily Means records \citep{dataset}. SORCE carried four instruments: the Total Irradiance Monitor (TIM), the Spectral Irradiance Monitor (SIM), the Solar Stellar Irradiance Comparison Experiment (SOLSTICE), and the X-Ray/XUV Photometer System (XPS) \citep{rottman2005sorce,woods2021overview}. The XPS instrument used broadband photometers with bandpass filters to monitor solar X-ray ultraviolet irradiance shortward of approximately 40 nm and the H I Lyman-$\alpha$ emission at 121.6 nm \citep{woods2005xps_overview,woods2022xpsfinal}.

Solar spectral irradiance (SSI) is the solar radiative power received per unit
area and per unit wavelength at a given distance from the Sun
\citep{foukal2008solar}. For a finite spectral band, the reported irradiance
corresponds to the integral of SSI over that wavelength interval. Therefore, the
series analyzed here is the band integrated SSI measured by the selected
SORCE/XPS photodiode, rather than the full wavelength resolved SSI spectrum \citep{woods2021overview}.

We analyze photodiode 7, which monitors the 0.1--7.0 nm spectral band. The selected data product reports observation date, photodiode identifier, median irradiance, mean irradiance, measurement precision, calculation precision, and absolute uncertainty. Irradiance is reported at a mean Sun-Earth distance of 1 AU and with line-of-sight relative velocity set to zero with respect to the Sun \citep{dataset}. The uncertainty and precision quantities are reported at the $1\sigma$ level.

The analysis uses the mean solar spectral irradiance (SSI) series in W m$^{-2}$ from 1 January 2005 to 31 December 2019. Although SORCE operated over a longer interval, this 15-year window was selected to obtain a consistent analytical period with adequate data availability and a regular structure suitable for historical series and EVT procedures. The endpoint was set before the termination of the mission in February 2020. The original observations are not perfectly aligned with a complete daily grid. We therefore apply linear interpolation to construct the daily analytical series used in the subsequent analyses. Interpolated values are treated as estimated values on a regular temporal grid, not as additional physical measurements.

SORCE mission operations provide important context for the statistical analysis. The mission produced daily averaged solar irradiances for most of its lifetime, but it also experienced data gaps, including the largest gap from August 2013 to February 2014 associated with declining battery capacity and the later restriction of operations to daylight conditions \citep{woods2021overview}. Such gaps matter for tail inference because extreme value methods depend strongly on the timing, magnitude, and availability of the most unusual observations.

\subsection{Measurement model and reported uncertainty}

Let \(t\) denote the time index over the observational period, restricted here to
the ordered records at which SORCE/XPS observations are available. Thus,
\(t=1,\ldots,n_{\mathrm{obs}}\) indexes the observed records. Let \(X_t^\ast\)
denote the latent solar spectral irradiance at time \(t\), that is, the
unobserved physical quantity that the instrument is intended to measure. The
irradiance reported by the SORCE/XPS processing pipeline at time \(t\) is denoted by \(X_t^{\mathrm{obs}}\) .

Following the usual measurement error representation \citep{fuller1987measurement},
we use the conceptual model
\begin{equation}
X_t^{\mathrm{obs}}
=
X_t^{\ast}
+
\varepsilon_t,
\qquad t = 1,\ldots,n_{\mathrm{obs}},
\label{eq:measurement_model}
\end{equation}
where \(\varepsilon_t\) represents the combined contribution of measurement
error, calibration uncertainty, processing uncertainty, and other effects associated
with the instrument and data pipeline. No parametric distribution is assumed for
\(\varepsilon_t\), and the present analysis does not attempt to estimate this
error term separately.

The SORCE/XPS reports measurement precision, calculation precision,
and absolute uncertainty. In line with standard guidance on the expression of
measurement uncertainty \citep{jcgm2008gum}, these quantities are not modeled as additional response
variables in the present univariate analysis. Rather, they are used to characterize
the quality of the measurements and to guide the interpretation of tail estimates
obtained from the processed instrumental record. When absolute uncertainty is
reported in W m\(^{-2}\), it is interpreted as uncertainty attached to the reported
irradiance value, not as an additional observation. 
We therefore do not combine measurement precision, calculation precision, and absolute uncertainty into a
single variance unless their interpretation as separate uncertainty components is
explicitly established by the data documentation.

From the perspective of systematic-error analysis, the contribution of the present study lies in keeping the observation process explicit when interpreting tail estimates, rather than treating the processed irradiance
series as an error-free realization of the latent solar signal.

\subsection{Interpolation and analytical time grid}

To apply probability models to the SORCE/XPS record, such as the extreme value distributions, it is necessary to address the presence of missing observations. Therefore, this section describes the interpolation technique proposed by \citep{hamilton2020time} to construct a complete daily time series by estimating the missing values, while linear interpolation provides a simple and transparent way to estimate values between neighboring observations \citep{lepot2017interpolation}.

Let \(t=1,\ldots,n_{\mathrm{obs}}\) index the ordered records at which
SORCE/XPS observations are available in the original record. These observations
are not necessarily available on consecutive days. Let \(s\) denote a day in the complete daily grid used to construct the analytical series.

Suppose that the observation corresponding to day $s$ is missing, while the observations before and after this day are available. The missing value can then be estimated using the interpolation procedure described below.

For any day \(s\)
such that
\[
a \leq s \leq b,
\]
where \(a\) and \(b\) are two adjacent observed days in the original record, the
daily value is defined by linear interpolation as
\begin{equation}
\widetilde{X}_{s}
=
(1-w_s)X_{a}^{\mathrm{obs}}
+
w_s X_{b}^{\mathrm{obs}},
\qquad
w_s
=
\frac{s-a}{b-a}.
\label{eq:interpolation}
\end{equation}

In this context, missing values inside an observed interval are replaced by values along the linear segment connecting the adjacent observations; they are not treated as constant, repeated, or newly measured irradiance values.

After interpolation, the complete daily analytical series is reindexed as \(\{\widetilde{X}_t\}_{t=1}^{n}\), where \(t\) denotes the position of a day in the regular analytical sequence. Interpolated values are treated as derived estimates on a regular temporal grid, not as additional physical measurements. For this reason, the interpolation step is considered part of the construction of the analytical series, whereas the EVT models are fitted only after this regular grid has been defined. Because linear interpolation cannot recover unobserved short time scale variability within missing data, the resulting daily series should be interpreted as a regular analytical representation of the available record, not as a reconstruction of all physical fluctuations during missing intervals.

\subsection{Daily logarithmic changes in irradiance and inferential target}

Because irradiance is positive and because our inferential interest is the relative variation from one day to the next, we analyze daily logarithmic changes rather than raw differences. This transformation expresses changes between consecutive observations as logarithmic ratios and provides a dimensionless representation of day-to-day fluctuations.

For the daily interpolated SSI series
\(\{\widetilde{X}_t\}_{t=1}^{n}\), the logarithmic change is defined as

\begin{equation}
r_t
=
\log(\widetilde{X}_t)
-
\log(\widetilde{X}_{t-1})
=
\log\left(
\frac{\widetilde{X}_t}{\widetilde{X}_{t-1}}
\right),
\qquad
t=2,\ldots,n.
\label{eq:daily_log_change}
\end{equation}

This transformation is useful in the present setting because it expresses
changes in irradiance relative to the preceding day. Thus a positive value of
\(r_t\) represents a relative increase in the reported SSI, whereas a negative
value represents a relative decrease. The corresponding proportional change is

\begin{equation}
\exp(r_t)-1
=
\frac{\widetilde{X}_t-\widetilde{X}_{t-1}}
{\widetilde{X}_{t-1}},
\label{eq:relative_change_interpretation}
\end{equation}

\noindent and, for small changes, \(\exp(r_t)-1\approx r_t\). This approximation is used
only for interpretation; all subsequent statistical analyses are performed on
the logarithmic-change scale.

The reported measurement precision, calculation precision, and absolute
uncertainty are not propagated analytically into \(r_t\) in this study.
Instead, they are used to contextualize the quality of the SORCE/XPS record
and to guide the interpretation of tail estimates obtained from the processed
instrumental series.

The inferential target is the tail behavior of the series of daily logarithmic
irradiance changes \(\{r_t\}\). Positive extremes are modeled through block
maxima of \(r_t\), whereas negative extremes are modeled through transformed
block minima. This allows both tails to be analyzed using the standard maximum
form of the GEV distribution \citep{kotz2000extreme}.

\subsection{Block maxima and transformed minima}

The series of daily logarithmic irradiance changes \(\{r_t\}_{t=2}^{n}\) was divided into non-overlapping blocks following the standard block approach in Extreme Value Theory \citep{embrechts1997modelling,kotz2000extreme}. For a fixed candidate block length \(q\), let \(B_j\) denote the set of time indices contained in block \(j\), where \(j=1,\ldots,k\) and \(k\) is the number of complete blocks. The upper and lower tail extremes were defined as

\begin{equation}
M_j^{+}
=
\max_{t\in B_j} r_t,
\qquad
M_j^{-}
=
-\min_{t\in B_j} r_t.
\label{eq:block_extremes}
\end{equation}

The sequence \(\{M_j^{+}\}\) contains the largest daily logarithmic irradiance change within each block and therefore represents extreme relative increases in irradiance. The sequence \(\{M_j^{-}\}\) contains the magnitude of the most negative daily log-change within each block and represents extreme relative reductions. The sign
transformation allows both sequences to be modeled using the standard maximum
formulation of the GEV distribution. When interpreting the lower tail on the original log-change scale, \(M_j^{-}\) corresponds to the negative log-change
\(-M_j^{-}\).

Block lengths were selected separately for the two tails using the
Ljung-Box test \citep{ljung1978} and visual inspection of the sample
autocorrelation function \citep{box2015time}. For a sequence of \(k\) block
extremes, the Ljung-Box statistic at lag \(h\) is

\begin{equation}
Q(h)
=
k(k+2)
\sum_{\ell=1}^{h}
\frac{\widehat{\rho}_{\ell}^{\,2}}{k-\ell},
\label{eq:ljung_box}
\end{equation}

where \(\widehat{\rho}_{\ell}\) is the sample autocorrelation at lag
\(\ell\). The null hypothesis is that the autocorrelations up to lag \(h\)
are equal to zero. The test was therefore used as a diagnostic of residual
serial dependence rather than as proof of complete independence.

Candidate block lengths were screened at a significance level of
\(\alpha_{\mathrm{LB}}=0.025\). This threshold was adopted as a pragmatic
criterion for balancing weak residual dependence against the need to retain a
sufficient number of block extremes in a finite instrumental record affected
by data gaps and interpolation. Selection was not based on the Ljung-Box
\(p\)-value alone: the corresponding autocorrelation functions and the
resulting effective sample sizes were also examined. The selected block
lengths and numbers of retained extremes are reported in
Section~\ref{subsec:results_blocks}.

\subsection{Generalized Extreme Value distribution and parameter estimation}

Let \(Y_1,\ldots,Y_k\) denote the block maxima or transformed block minima
retained for a given fit, where \(k\) is the number of complete blocks. Let
\(y_1,\ldots,y_k\) denote their observed values. The transformed minima were defined as the negative of the minimum daily logarithmic irradiance change within each block, allowing both tails to be modeled using the standard maximum formulation.

Under standard extreme value conditions, block extremes can be approximated by
the Generalized Extreme Value distribution
\citep{fisher1928limiting,gnedenko1943,von1936distribution,
jenkinson1955frequency, decarvalho2026handbook}. Its cumulative distribution function is

\begin{equation}
G_Y(y;\mu,\sigma,\xi)
=
\begin{cases}
\exp\left\{
-\left[
1+\xi\left(\dfrac{y-\mu}{\sigma}\right)
\right]^{-1/\xi}
\right\},
& \xi\neq0, \\[1em]
\exp\left\{
-\exp\left[-\left(\dfrac{y-\mu}{\sigma}\right)\right]
\right\},
& \xi=0,
\end{cases}
\label{eq:gev}
\end{equation}

where \(\mu\in\mathbb{R}\) is the location parameter,
\(\sigma>0\) is the scale parameter, and \(\xi\in\mathbb{R}\) is the shape
parameter. For \(\xi\neq0\), the support satisfies

\begin{equation}
1+\xi\left(\frac{y-\mu}{\sigma}\right)>0.
\label{eq:gev_support}
\end{equation}

The cases \(\xi>0\), \(\xi=0\), and \(\xi<0\) correspond respectively to
Fr\'echet, Gumbel, and Weibull type upper tails
\citep{kotz2000extreme,embrechts1997modelling}.

The parameters were primarily estimated by maximum likelihood. For the observed
block extremes \(y_1,\ldots,y_k\), the GEV log likelihood is

\begingroup
\small
\begin{equation}
\ell(\mu,\sigma,\xi)
=
\left\{
\begin{aligned}
&-k\log\sigma
-\sum_{i=1}^{k}
\left(\frac{y_i-\mu}{\sigma}\right)
\\[-0.1em]
&\qquad
-\sum_{i=1}^{k}
\exp\left[
-\left(\frac{y_i-\mu}{\sigma}\right)
\right],
&& \xi=0,
\\[0.8em]
&-k\log\sigma
-\left(1+\frac{1}{\xi}\right)
\sum_{i=1}^{k}
\log\left[
1+\xi\left(\frac{y_i-\mu}{\sigma}\right)
\right]
\\[-0.1em]
&\qquad
-\sum_{i=1}^{k}
\left[
1+\xi\left(\frac{y_i-\mu}{\sigma}\right)
\right]^{-1/\xi},
&& \xi\neq0.
\end{aligned}
\right.
\label{eq:gev_loglik}
\end{equation}
\endgroup

For \(\xi\neq0\), maximization is subject to

\begin{equation}
\sigma>0
\qquad\text{and}\qquad
1+\xi\left(\frac{y_i-\mu}{\sigma}\right)>0,
\quad i=1,\ldots,k.
\label{eq:gev_likelihood_support}
\end{equation}

The maximum likelihood estimator is

\begin{equation}
(\widehat{\mu},\widehat{\sigma},\widehat{\xi})_{\mathrm{MLE}}
=
\argmax_{\mu,\sigma,\xi}
\ell(\mu,\sigma,\xi).
\label{eq:gev_mle}
\end{equation}

Probability-weighted moments were also used as a complementary estimation
method \citep{greenwood1979probability,hosking1985estimation}. PWM provides a
moment based alternative to maximum likelihood and was included to assess the
sensitivity of the GEV parameter estimates, particularly given the limited
number of block extremes available for some fits. Both methods estimate the
same parameter vector \((\mu,\sigma,\xi)\). Maximum likelihood estimates were
used as the primary inferential reference, whereas PWM estimates were used only
for comparison and sensitivity assessment.

\subsection{Return levels}

Return levels are quantiles of the block-extreme distribution associated with
specified mean recurrence intervals \citep{kotz2000extreme,embrechts1997modelling}.

Let \(q\) denote the block length in days and let $ \nu_q= 365.25/q$ be the approximate number of non-overlapping blocks per year. The return level
\(z_{T,q}\) associated with a return period of \(T\) years is defined by

\begin{equation}
\Prob(Y>z_{T,q})
=
\frac{1}{\nu_qT},
\label{eq:return_definition}
\end{equation}

The corresponding GEV quantile is

\begin{equation}
z_{T,q}
=
\begin{cases}
\begin{aligned}
&\mu
-\sigma
\log\left[
-\log\left(1-\frac{1}{\nu_qT}\right)
\right],
\end{aligned}
& \xi=0, \\[1.5em]
\begin{aligned}
&\mu+\frac{\sigma}{\xi}
\left[
\left\{
-\log\left(1-\frac{1}{\nu_qT}\right)
\right\}^{-\xi}
-1
\right],
\end{aligned}
& \xi\neq0.
\end{cases}
\label{eq:returnlevel}
\end{equation}

Estimated return levels were obtained by replacing
\((\mu,\sigma,\xi)\) with their maximum likelihood estimates \eqref{eq:gev_mle}. Return periods represent mean recurrence intervals under the fitted stationary model and
should not be interpreted as deterministic waiting times. Uncertainty in the
estimated GEV parameters propagates directly to the return levels and becomes
increasingly relevant for return periods extending beyond the observed record.

Because the models were fitted to logarithmic changes, an upper-tail return level \(z_{T,q}\) corresponds to the relative increase $ \exp(z_{T,q})-1$. For the transformed block minima, the corresponding signed logarithmic change
is \(-z_{T,q}\), which represents the relative reduction
\(1-\exp(-z_{T,q})\).

\subsection{Software and reproducibility}
\label{subsec:software}

All analyses were conducted in R version 4.5.1 \citep{R_base}. The computational workflow included time series preprocessing, block extreme extraction, GEV fitting, model diagnostics, and return level calculation. Maximum likelihood GEV estimates and return levels were obtained with the \texttt{extRemes} package \citep{gilleland2016extremes}, whereas estimates based on probability weighted moments, used as a sensitivity analysis, were obtained with the \texttt{fExtremes} package \citep{wuertz2009package}.

Pointwise 95\% confidence bands for the return levels were obtained using a parametric bootstrap procedure. For each fitted tail, samples were simulated from the maximum likelihood GEV model, the GEV parameters \((\mu,\sigma,\xi)\) were re-estimated for each bootstrap sample, and the
corresponding return level curves were calculated. The pointwise 2.5th and 97.5th percentiles of the bootstrap return levels were used as the lower and upper confidence limits, respectively. These bands quantify parameter estimation uncertainty conditional on the fitted stationary GEV model, the selected block construction, and the processed analytical series. They do not incorporate measurement or interpolation uncertainty, or uncertainty arising from possible model misspecification.

All numerical results, tables, and figures were obtained from the same final
daily interpolated series.

\section{Results}
\label{sec:results}

\subsection{Temporal and marginal behavior of the SORCE/XPS irradiance record}
\label{subsec:results_descriptive}

\begin{figure*}[!t]
    \centering
    \includegraphics[width=0.8\textwidth]{\detokenize{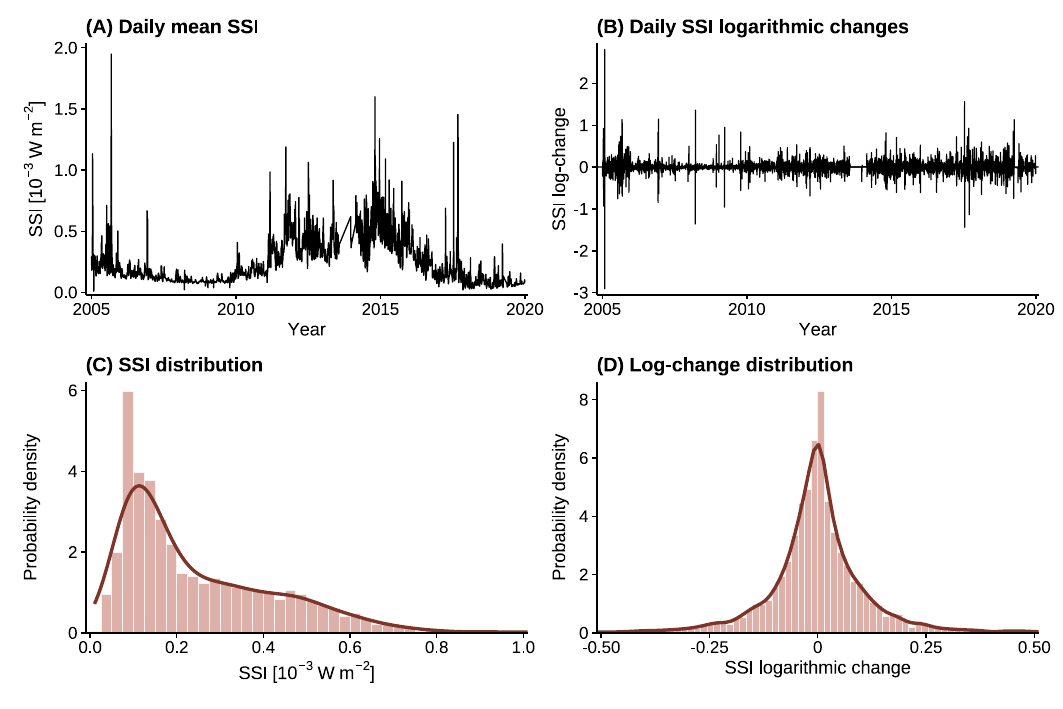}}
    \caption{SORCE/XPS solar spectral irradiance (SSI) and daily logarithmic
    changes from 2005 to 2019.
    (A) Daily SSI measured by photodiode~7 and integrated over 0.1 to 7.0~nm on
    the regular analytical grid, in \(\mathrm{W\,m^{-2}}\).
    (B) Daily logarithmic irradiance changes.
    (C) Empirical distribution of SSI.
    (D) Empirical distribution of the daily logarithmic changes.}
    \label{figure_1}
    
    \vspace{0pt}
    
    \includegraphics[width=0.7\textwidth]{\detokenize{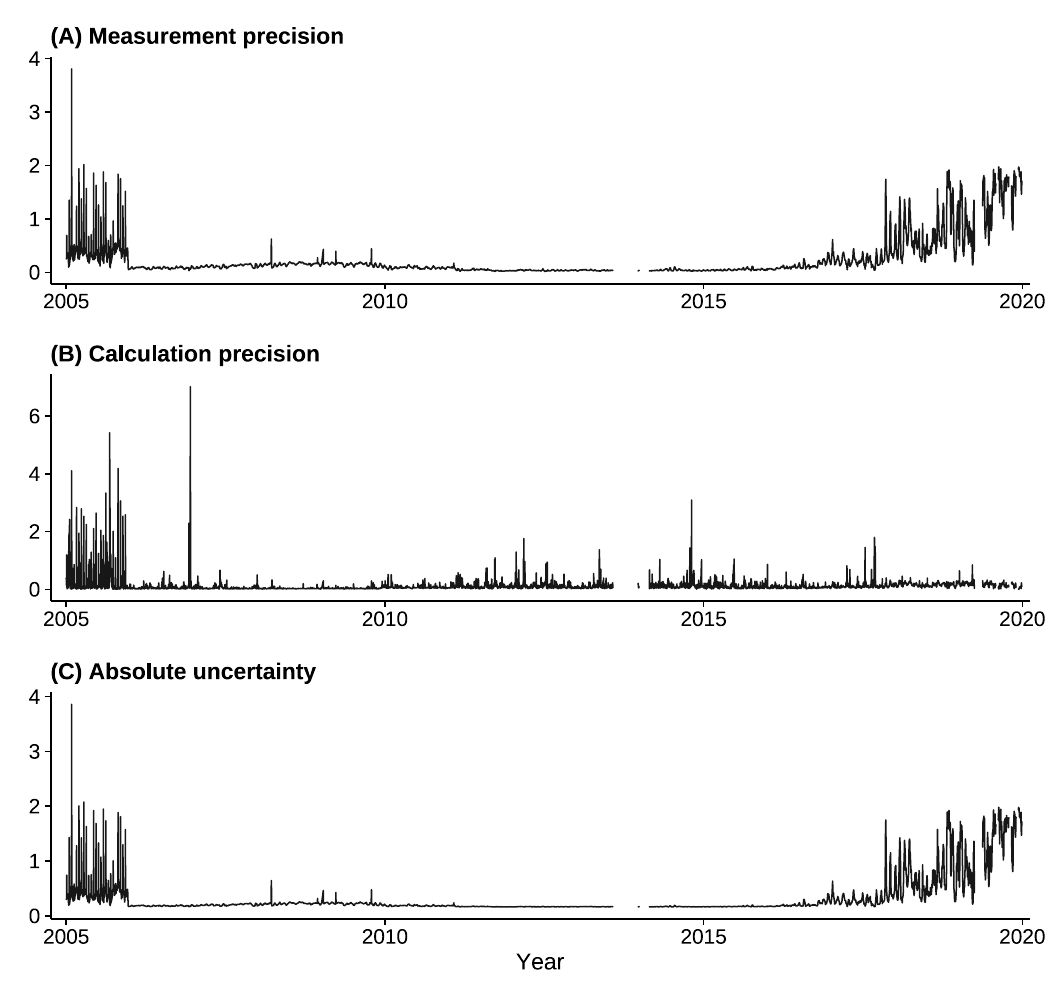}}
    \caption{Reported precision and uncertainty measures for SORCE/XPS
    photodiode 7 (0.1 to 7.0~nm).
    (A) Measurement precision.
    (B) Calculation precision.
    (C) Absolute uncertainty.
    All quantities are associated with the corresponding Level~3 daily mean
    irradiance records and are reported at the \(1\sigma\) level.}
    \label{figure_2}
\end{figure*}

Figure~\ref{figure_1} summarizes the final daily analytical record for
SORCE/XPS photodiode~7, which measures solar irradiance obtained by integrating
SSI over the spectral interval from 0.1 to 7.0~nm. On the irradiance scale, the distribution is strongly
right-skewed: the sample mean (\(0.262\times10^{-3}\ {\rm W\,m^{-2}}\)) exceeds the median
(\(0.2060\times10^{-3}\ {\rm W\,m^{-2}}\)), the sample skewness is 1.626, and
the maximum (\(1.9478\times10^{-3}\ {\rm W\,m^{-2}}\)) is approximately one
order of magnitude larger than the median
(Table~\ref{tab:descriptive}). The time series also shows pronounced
temporal heterogeneity, with a broad interval of increased irradiance and larger
fluctuations during the middle and later portions of the record. This behavior is consistent with the strong variability expected in solar emission at short wavelengths across different levels of solar activity
\citep{woods2005xps_variations,woods2021overview}; however the present analysis
does not assign individual observations to specific flares, active regions, or
other physical events.

\begin{table}[!t]
\centering
\caption{Descriptive statistics of the daily analytical SORCE/XPS band-integrated mean solar spectral irradiance measured by photodiode~7 over
0.1--7.0~nm. Irradiance statistics, except skewness, are reported in \(10^{-3}\,\mathrm{W\,m^{-2}}\).}
\label{tab:descriptive}
\begin{tabular}{lr}
\toprule
Statistic & Value \\
\midrule
Minimum            & 0.0100 \\
1st quartile        & 0.1126 \\
Median              & 0.2060 \\
Mean                & 0.2625 \\
3rd quartile        & 0.3747 \\
Maximum             & 1.9478 \\
Standard deviation & 0.1897 \\
Skewness            & 1.6261 \\
\bottomrule
\end{tabular}
\end{table}

The series of daily logarithmic changes \(r_t\), defined in
Eq.~\eqref{eq:daily_log_change}, is sharply concentrated near zero but contains
observations extending into both tails. Its interpretation follows
Eq.~\eqref{eq:relative_change_interpretation}: a value \(r_t>0\) corresponds
to a relative increase \(\exp(r_t)-1\), whereas \(r_t<0\) corresponds to a
relative reduction \(1-\exp(r_t)\) with respect to the preceding daily value.
Thus, the central concentration of daily changes coexists with a small number
of unusually large relative increases and reductions, motivating separate
analyses of the two tails.

Figure~\ref{figure_2} displays the measurement precision, calculation
precision, and absolute uncertainty reported for the photodiode at the
\(1\sigma\) level. These quality indicators vary over the mission record rather
than remaining constant in time. Their temporal variation does not identify a
specific error mechanism by itself, but it shows that the reliability of the
reported irradiance is not homogeneous throughout the observational period.
Accordingly, the extremes analyzed below are extremes of the processed
instrumental record \(X_t^{\mathrm{obs}}\), not direct observations of the
latent irradiance \(X_t^\ast\) in Eq.~\eqref{eq:measurement_model}. The
reported uncertainty quantities are therefore used to qualify the tail
inference and are not interpreted as realizations of the measurement error
\(\varepsilon_t\).

\begin{figure*}[!t]
    \centering
    \includegraphics[width=0.8\textwidth]{\detokenize{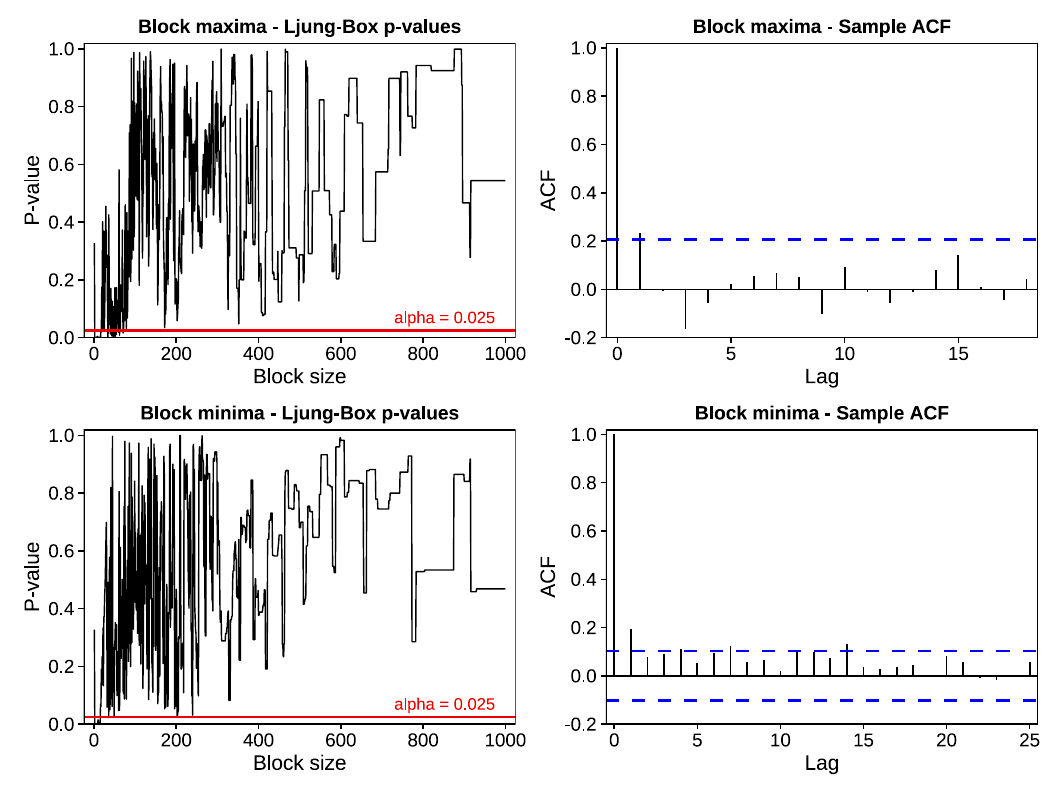}}
    \caption{Block length selection and residual dependence diagnostics for upper
    and lower tail extremes. Ljung-Box \(p\)-values are shown across candidate
    block lengths, together with the sample ACFs at the selected values. The final
    analysis uses blocks of 60 days for upper tail maxima and blocks of 15 days for
    transformed lower tail minima.}
    \label{figure_3}
\end{figure*}

\subsection{Block selection and temporal dependence diagnostics}
\label{subsec:results_blocks}

Figure~\ref{figure_3} summarizes the Ljung-Box screening results and the
autocorrelation functions used to evaluate the candidate block lengths. The
procedure described after Eq.~\eqref{eq:ljung_box} selected a block length of
\(q_{+}=60\) days for the upper tail and \(q_{-}=15\) days for the lower
tail. These choices yielded \(k_{+}=90\) block maxima \(M_j^{+}\) and
\(k_{-}=364\) transformed block minima \(M_j^{-}\), respectively.

The upper tail sequence showed weak residual autocorrelation at the selected
block length and was considered adequate for the subsequent GEV approximation.
The shorter lower tail construction preserved a larger number of extreme
observations, although some small residual autocorrelation features remained.
The two sequences were therefore treated as approximately independent for
modeling purposes, rather than as sequences for which complete stochastic
independence had been established.

The different selected block lengths indicate that extreme positive and
negative daily changes did not exhibit identical temporal dependence
structures. Comparisons between the fitted tails must therefore account for
their different aggregation scales and effective sample sizes.

\subsection{GEV estimates and model diagnostics}
\label{subsec:results_gev}

\begin{figure*}[!t]
    \centering
    \includegraphics[width=0.85\textwidth]{\detokenize{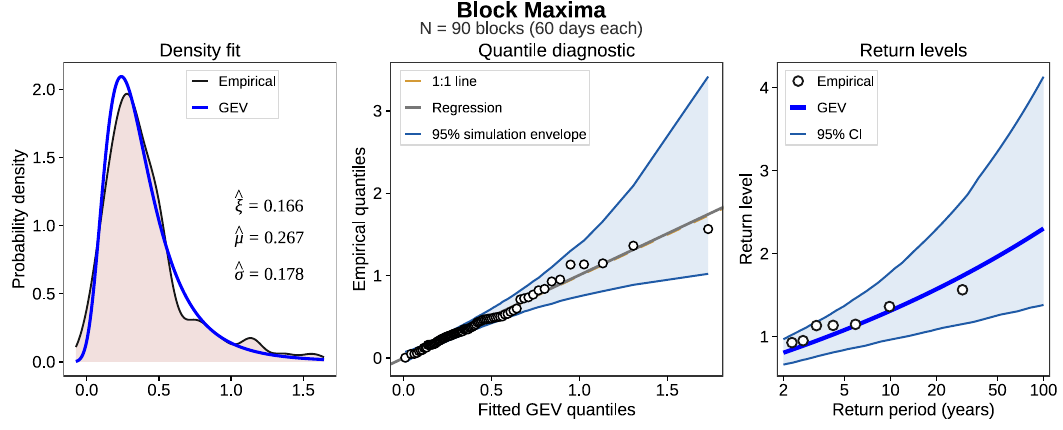}}
    \caption{GEV fit, quantile diagnostic, and annualized return levels for block maxima of the SORCE/XPS series of daily logarithmic changes. From left to right, the panels show the empirical and fitted GEV densities for maxima obtained from 60 day blocks, the comparison between empirical and fitted quantiles with the 1:1 reference line, regression line, and 95\% simulation envelope, and the annualized return level estimates with the fitted curve and 95\% bootstrap confidence bands as functions of the return period.}
    \label{figure_4}
\end{figure*}

\begin{figure*}[!t]
    \centering
    \includegraphics[width=0.85\textwidth]{\detokenize{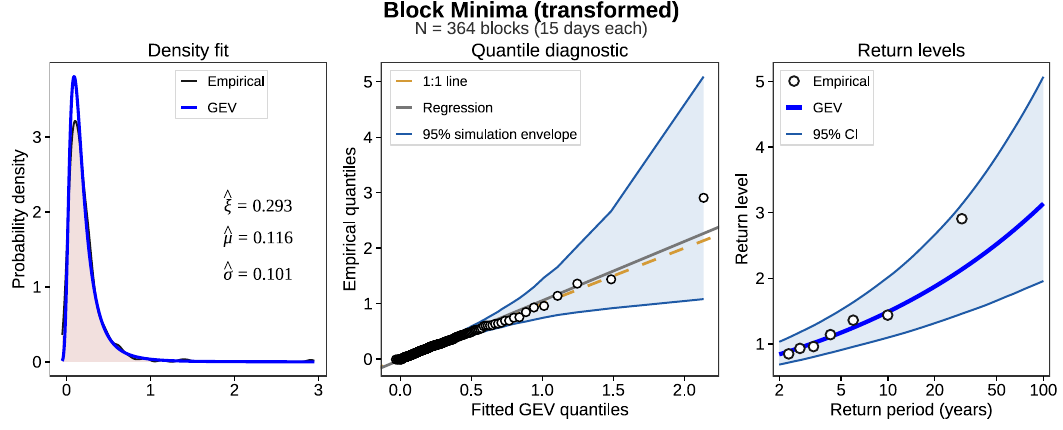}}
    \caption{GEV fit, quantile diagnostic, and annualized return levels for
    transformed block minima of the SORCE/XPS series of daily logarithmic changes. From left to right, the panels show the empirical and fitted GEV densities for the transformed minima, obtained from 15 day blocks, the comparison between empirical and fitted quantiles with the 1:1 reference line, regression line, and 95\% simulation envelope, and the annualized return level estimates with the fitted curve and 95\% bootstrap confidence bands as functions of the return period.}
    \label{figure_5}
\end{figure*}

The GEV model in Eq.~\eqref{eq:gev} was fitted separately to the upper tail
block maxima \(\{M_j^{+}\}\) and the transformed lower tail block minima
\(\{M_j^{-}\}\). Maximum likelihood estimation was used as the primary
inferential method, whereas probability weighted moments were used to assess
the sensitivity of the estimates to the estimation procedure. The complete
parameter estimates are reported in Table~\ref{tab:gev_parameters} in the
standardized order \((\mu,\sigma,\xi)\).

\begin{table}[!b]
\centering
\caption{GEV parameter estimates for upper tail block maxima and transformed
lower tail block minima. Here, \(q\) is the block length in days and \(k\) is
the number of complete blocks; MLE is the primary estimator and PWM is included
for sensitivity analysis.}
\label{tab:gev_parameters}
\footnotesize
\setlength{\tabcolsep}{2.8pt}
\renewcommand{\arraystretch}{1.08}
\begin{tabularx}{\columnwidth}{@{}L l r r c c c@{}}
\toprule
Extreme series & Method & \(q\) & \(k\) &
\(\widehat{\mu}\) & \(\widehat{\sigma}\) & \(\widehat{\xi}\) \\
\midrule
Block maxima             & MLE & 60 & 90  & 0.2668 & 0.1780 & 0.1661 \\
Block maxima             & PWM & 60 & 90  & 0.2641 & 0.1750 & 0.1848 \\
\midrule
Block minima (transformed) & MLE & 15 & 364 & 0.1157 & 0.1005 & 0.2928 \\
Block minima (transformed) & PWM & 15 & 364 & 0.1160 & 0.1020 & 0.2830 \\
\bottomrule
\end{tabularx}
\end{table}

MLE and PWM produced closely aligned estimates for both fitted tails. For the
60 day upper tail blocks, the shape estimates were
\(\widehat{\xi}_{+}=0.1661\) under MLE and
\(\widehat{\xi}_{+}=0.1848\) under PWM. For the transformed 15 day lower tail
blocks, the corresponding estimates were
\(\widehat{\xi}_{-}=0.2928\) and
\(\widehat{\xi}_{-}=0.2830\). Thus, the principal conclusions about tail form
were not materially altered by the choice between the two estimation methods.

At their point estimates, all fitted shape parameters are positive and are
therefore compatible with Fr\'echet type tails under the adopted block
constructions. For \(M_j^{-}\), this result refers to the upper tail of the
transformed magnitudes and corresponds, after reversing the sign, to increasingly negative log-changes in SSI. Nevertheless the positive estimates alone do not establish that \(\xi>0\) with statistical significance or exclude the Gumbel
case \(\xi=0\); such conclusions would require confidence intervals or a
profile likelihood analysis for the shape parameter.

The larger point estimates obtained for the transformed lower tail provide
descriptive evidence that the magnitude of extreme reductions may have a
heavier fitted tail than that of extreme increases. This comparison should not,
however, be interpreted as a formal test of tail asymmetry because the models
were fitted using different block lengths and substantially different numbers
of block extremes.

The diagnostic plots in Figures~\ref{figure_4} and~\ref{figure_5} show that the
fitted distributions reproduce the main structure of both extreme samples and
that most empirical quantiles are reasonably consistent with their fitted
counterparts. The most visible departures occur among the largest observations,
where the amount of information is necessarily limited. The GEV models are
therefore adequate as parsimonious descriptions of the observed block extremes,
although inference in the farthest tails remains sensitive to the shape
parameter and the upper order statistics
\citep{kotz2000extreme,embrechts1997modelling}.

\subsection{Annualized return levels and relative changes}
\label{subsec:results_returnlevels}

Block-based return levels were indexed by annual return periods according to
Eq.~\eqref{eq:return_definition} and Eq.~\eqref{eq:returnlevel}, using
\(\nu_q=365.25/q\), where \(q\) is the block length in days. Because the two
tails were modeled with different block lengths, the upper tail levels are
denoted by \(z^{+}_{T,60}\), whereas the transformed lower tail levels are
denoted by \(z^{-}_{T,15}\). The MLE point estimates for all considered return
periods are reported in Table~\ref{tab:return_levels}.

\begin{table}[!b]
\centering
\caption{Annualized GEV return levels and corresponding relative changes for
block maxima (60 days) and transformed block minima (15 days). Return levels are expressed on the logarithmic-change scale and are based on the stationary MLE fits.}
\label{tab:return_levels}
\footnotesize
\setlength{\tabcolsep}{2.4pt}
\renewcommand{\arraystretch}{1.08}
\begin{tabularx}{\columnwidth}{@{}cYYYY@{}}
\toprule
& \multicolumn{2}{c}{Block maxima}
& \multicolumn{2}{c}{Block minima (transformed)}\\
\cmidrule(lr){2-3}\cmidrule(lr){4-5}
\(T\) (years) &
\(\widehat{z}^{+}_{T,60}\) &
Increase (\%) &
\(\widehat{z}^{-}_{T,15}\) &
Reduction (\%) \\
\midrule
2   & 0.8068 & 124.1 & 0.8401 & 56.8 \\
5   & 1.0798 & 194.4 & 1.1711 & 69.0 \\
20  & 1.5728 & 382.0 & 1.8733 & 84.6 \\
50  & 1.9648 & 613.3 & 2.5203 & 92.0 \\
100 & 2.3031 & 900.5 & 3.1387 & 95.7 \\
\bottomrule
\end{tabularx}
\end{table}

For the upper tail, the estimated return level increases from
\(\widehat{z}^{+}_{2,60}=0.8068\) at a two year return period to
\(\widehat{z}^{+}_{100,60}=2.3031\) at a 100 year return period. On the relative
change scale, these estimates correspond to increases of approximately
\(124.1\%\) and \(900.5\%\), respectively, relative to the irradiance observed
on the preceding day.

For the transformed lower tail, the estimated level increases from
\(\widehat{z}^{-}_{2,15}=0.8401\) to
\(\widehat{z}^{-}_{100,15}=3.1387\). On the original log-change scale, the corresponding signed logarithmic changes are \(-\widehat{z}^{-}_{T,15}\). Applying the transformation \(1-\exp(-\widehat{z}^{-}_{T,15})\), the two estimates represent
relative reductions of approximately \(56.8\%\) and \(95.7\%\).

These percentages describe changes relative to the immediately preceding daily
solar spectral irradiance value. They are not return levels of absolute SSI in
\({\rm W\,m^{-2}}\), estimates of instantaneous flare intensity, or forecasts
of specific solar events. In addition, the percentages for the two tails should
not be compared directly as measures of physical asymmetry. Under the
logarithmic transformation, relative increases are unbounded, whereas relative
reductions are bounded above by \(100\%\).

The return levels increase with \(T\). The positive fitted shape parameters
indicate heavy, unbounded upper tails on the respective modeled scales, leading
to increasingly pronounced extrapolation at long return periods. However, the
50 and 100 year estimates extend far beyond the 15 year observational record.
They should therefore be interpreted as conditional model extrapolations whose
reliability depends on the assumed stationarity of the GEV models, the selected
block constructions, residual temporal dependence, interpolation of the daily
series, and uncertainty in \((\mu,\sigma,\xi)\). Return periods represent mean
recurrence intervals under the fitted models and not deterministic waiting
times.

\section{Discussion}
\label{sec:discussion}

\subsection{Differences between positive and negative extreme changes}
\label{subsec:discussion_asymmetry}

Different block lengths were required for the two tails to obtain sequences with practically acceptable residual dependence. The upper-tail analysis retained 60-day block maxima, whereas the lower-tail analysis retained transformed minima from 15-day blocks. This result suggests that the dependence structure associated with large positive changes differs from that associated with large negative changes. It should not, however, be interpreted as establishing fixed physical clustering timescales for either type of event.
 
The fitted shape point estimate was larger for the transformed minima than for the maxima. Conditional on the selected preprocessing and block definitions, this is compatible with a more pronounced lower tail for the daily logarithmic irradiance changes. The apparent difference in tail heaviness should nevertheless be interpreted with caution, since the two shape estimates were obtained using different block lengths and different numbers of extreme observations.

Interpretation must also account for the relative nature of daily logarithmic change. A large negative log-change may follow an unusually elevated daily solar spectral irradiance value, whereas a large positive log-change may represent a rapid increase from a comparatively low baseline. Consequently, the estimated difference between the two tails refers to relative daily changes and cannot be translated directly into an asymmetry of absolute radiative energy. Such a translation would require a model for the joint behavior of irradiance levels and their subsequent changes.

\subsection{Physical interpretation in the 0.1 to 7.0 nm band}
\label{subsec:discussion_physical}

The SORCE/XPS channel analyzed in this study measures irradiance integrated
over the 0.1 to 7.0~nm spectral band. Radiation in this wavelength range is
associated primarily with hot, optically thin plasma in the outer solar
atmosphere and is sensitive to coronal temperature structure, magnetic
activity, active region evolution, and enhancements associated with solar
flares
\citep{woods2005xps_overview,woods2005xps_variations,
aschwanden2006physics,del2018solar}. Some upper tail observations may therefore
reflect rapid increases in coronal emission. Lower tail observations may
include the decay following enhanced emission, changes in active region
visibility, rotational modulation, and other reductions in the band integrated
signal.

The present analysis does not separate the physical and instrumental processes
contributing to the Level 3 XPS record. Because observations are not necessarily
available daily, the series was placed on a regular daily grid during
preprocessing. The modeled extremes therefore represent unusual changes between
consecutive values on this analytical grid and may reflect solar variability,
temporal sampling, averaging, calibration, and data processing. They should not
be interpreted as instantaneous flare peaks, energies integrated over individual
flares, spectra of specific events, or probabilities of individual space weather
events.

Associating the fitted tails with specific solar phenomena would require
observations at a higher temporal cadence, spectral information, event
catalogues, and an explicit association procedure. Within the scope of this
study, the EVT analysis provides a probabilistic description of the magnitude
and recurrence of unusual changes in the processed 0.1 to 7.0~nm record. After
validation using independent data, these summaries may support comparisons
among observing periods or instruments and the statistical identification of
unusually large variations in short wavelength irradiance.

\subsection{Interpretation of return levels}
\label{subsec:discussion_returnlevels}

The estimated return levels provide conditional summaries of rare relative
changes in the processed SORCE/XPS record. Their interpretation depends on the
stationary GEV models defined in Eq.~\eqref{eq:return_definition} and
Eq.~\eqref{eq:returnlevel}. They therefore describe mean recurrence under the
fitted distributions rather than fixed waiting times or deterministic
predictions of future solar behavior.

Extrapolation becomes increasingly uncertain as the return period extends
beyond the 15 year observational record. This limitation is especially relevant
for the longest periods considered, since their estimates depend strongly on
the fitted shape parameter, the selected block lengths, and the most extreme
observations. These levels should consequently be interpreted as model based
scenarios rather than recurrence patterns directly established by the data.

Stationarity also requires caution because the record includes different
phases of solar activity and changes in mission operation, while each tail is
represented by a single parameter vector throughout the study period. If the
distribution of extreme changes varies over time, the fitted return levels
represent averages across distinct observational and physical regimes.
Future analyses should quantify return level uncertainty, examine sensitivity
to block construction, and evaluate models whose parameters vary with
independent indicators of solar activity.

\subsection{Uncertainty in the processed record}
\label{subsec:discussion_measurement}

A central astrostatistical aspect of this study is the distinction between the
latent solar quantity \(X_t^\ast\) and the processed instrumental observation
\(X_t^{\mathrm{obs}}\), introduced in Eq.~\eqref{eq:measurement_model}. The
fitted tail models describe the processed observations rather than the latent
irradiance process. An extreme value in the analyzed record may therefore
reflect the solar signal together with detector response, calibration,
temporal sampling, averaging, and data processing.

The measurement/calculation precision and absolute uncertainty reported with the SORCE/XPS product provide complementary information about
the quality of the observed data \(X_t^{\mathrm{obs}}\). These quantities are not direct
realizations of the error term \(\varepsilon_t\) and should not be combined
into a single variance without a documented model for their definitions and
dependence structure
\citep{fuller1987measurement,jcgm2008gum}.

The reported \(1\sigma\) uncertainties were not propagated through the logarithmic-change transformation, block extraction, GEV estimation, or return level calculation. The resulting inference is therefore conditional on the processed analytical series and does not include all sources of uncertainty affecting the latent solar irradiance process. This distinction is particularly relevant for the shape parameter \(\xi\) and for long period return levels, which depend strongly on a limited number of extreme observations.

Linear interpolation introduces an additional interpretative consideration.
Equation~\eqref{eq:interpolation} provides a reproducible regular daily grid,
but it cannot reconstruct unobserved variability at shorter temporal scales.
Within data gaps, the interpolated path may smooth variations that were not
recorded and may influence returns near the boundaries of those gaps. This
issue is especially relevant to the interruption between August 2013 and
February 2014 documented for SORCE operations
\citep{woods2021overview}. The interpolated series should therefore be viewed
as an analytical representation of the available record rather than a complete
physical reconstruction of solar variability during periods without
observations.

Accordingly, the fitted tail estimates apply to the processed SORCE/XPS record
and identify the additional modeling required to infer extremes of the latent
irradiance process. Future work should examine the sensitivity of the results
to data gaps, reported uncertainty, and alternative treatments of the
observational record.

\subsection{Astrostatistical relevance and limitations}
\label{subsec:discussion_scope}

This study shows that interpretable extreme value inference can be obtained
from a processed space mission record even when the observational series is
affected by finite duration, incomplete temporal coverage, interpolation, and time varying measurement quality. By focusing on the tails of the daily log-change process, the analysis identifies features of the SORCE/XPS record that are not adequately represented by central summaries alone. Rare positive and negative changes have a substantial role in the fitted tail behavior and the return levels, which supports their explicit treatment in the statistical analysis of solar irradiance measurements.

The results remain conditional on the observational and inferential choices
adopted in this study. The record covers approximately 15 years, the upper tail fit is based on 90 block maxima, the block lengths were selected empirically, some residual dependence remains in the transformed minima sequence, and stationarity is assumed throughout. Measurement and interpolation uncertainty were not incorporated into the GEV likelihood. In addition, formal confidence intervals for the shape parameters and for the difference between the two tail indices are not currently available. The positive shape estimates should therefore be interpreted as compatible with Fr\'echet behavior under the selected block constructions, rather than as definitive evidence of an unbounded physical irradiance process.

These qualifications do not reduce the value of the fitted models; instead
they define the conditions under which the results are scientifically
interpretable. The analysis provides a coherent and reproducible connection
between the SORCE/XPS observations, the construction of a regular analytical series, relative irradiance changes, extreme events, GEV inference, and annualized return levels. Its main astrostatistical contribution is to show
that extreme behavior can be quantified without separating the statistical
model from the characteristics of the instrument, mission operations, and data processing.

The resulting models provide a baseline for comparing extreme behavior across
observing periods, instruments, and future solar missions. Further work should
propagate reported measurement uncertainty, examine sensitivity to data gaps and block construction, evaluate nonstationary models, and compare the inferred tails with independent observations or measurements at higher temporal
cadence. Multivariate analysis involving irradiance and instrumental
uncertainty represents a natural extension, provided that statistical
dependence is not naturally interpreted as physical causation.

\section{Conclusion}
\label{sec:conclusion}

This study developed a univariate probabilistic description of extreme daily changes in the band integrated 0.1 to 7.0~nm solar spectral irradiance measured by the SORCE/XPS mission from 2005 to 2019. The analysis of daily logarithmic changes placed relative increases and reductions on a common dimensionless scale. The fitted GEV models provided a consistent framework for characterizing unusually large positive and negative changes and for estimating annualized return levels. The positive shape parameter point estimates were compatible with Fr\'echet type tails under the selected block constructions, although this interpretation remains conditional on the finite record, block selection, and stationarity assumption.

Return levels give these results a temporal interpretation. Under the
stationary models adopted here, a level associated with a return period of
\(T\) years is expected to be exceeded, on average, once every \(T\) years.
This does not mean that such an event will occur at regular intervals. It is a
probabilistic statement about recurrence that allows rare fluctuations to be
compared on a common scale. The increasing uncertainty for long return periods also shows where inference from a 15 year record becomes most dependent on
statistical extrapolation.

The fitted models do not define physical limits for solar emission and should
not be viewed as a complete description of solar dynamics. They characterize a
finite instrumental record shaped by daily averaging, measurement uncertainty,
mission operations, data gaps, interpolation, block selection, residual
dependence, and the assumption of stationarity. Keeping these features visible
is essential because the most extreme observations are also those for which
measurement and preprocessing limitations may have the greatest inferential
consequences.

Within this scope, the study shows what astrostatistics can contribute to space
science: a transparent way to turn complex mission measurements into cautious
and interpretable statements about extreme behavior. This type of analysis can inform data quality assessment, uncertainty quantification, and statistical planning for future space missions. It also supports the broader study of extreme solar variability in the coupled Sun and Earth system, with relevance to satellites, communications, navigation, power infrastructure, upper atmospheric dynamics, and the investigation of meteorological and climate related processes. These connections matter because
the solar environment forms part of the broader physical context in which
atmospheric conditions, living systems, and essential human activities develop.
The contribution is therefore not a definitive model of solar extremes, but a
reproducible probabilistic framework that keeps statistical evidence,
instrumental limitations, and physical interpretation clearly separated.

\newpage
\section*{Conflict of Interest Statement}
The authors declare that the research was conducted in the absence of any commercial or financial relationships that could be construed as a potential conflict of interest.

\section*{Author Contributions}
EB contributed to the conceptualization, methodology, formal analysis, data visualization, and writing of the original draft. CEGO contributed to methodology, statistical overview, supervision, and critical revision of the manuscript. CB and BA contributed to the data processing, interpretation of the results and revision of the manuscript. FM contributed to the critical revision of the manuscript. All authors reviewed and approved the final version of the manuscript.

\section*{Funding}
The research itself did not receive specific project funding. During the period in which this work was developed, some of the authors received graduate fellowship support from the Brazilian National Council for Scientific and Technological Development (CNPq) and the Coordination for the Improvement of Higher Education Personnel (CAPES), Brazil.

\section*{Acknowledgments}
The authors acknowledge the SORCE mission team, NASA, the Laboratory for Atmospheric and Space Physics at the University of Colorado Boulder, and the Goddard Earth Sciences Data and Information Services Center for the development, processing, archiving, and public availability of the SORCE/XPS data products analyzed in this study. The authors also acknowledge the University of Bras\'ilia for financial support related to the presentation and dissemination of preliminary versions of this work at scientific meetings. EB received travel support to attend and present this work at the Systematic and Measurement Errors across the Sciences: AstroStatistics and Data Science (SYS2025) workshop through the Center for Space Plasma and Aeronomic Research at The University of Alabama in Huntsville, under U.S. National Science Foundation EPSCoR Award No. OIA-2505174.

\section*{Data Availability Statement}
The publicly available SORCE/XPS Level 3 dataset analyzed in this study is available through the NASA Goddard Earth Sciences Data and Information Services Center at \href{https://doi.org/10.5067/ZBKF34FGDCLT}{doi:10.5067/ZBKF34FGDCLT} \citep{dataset}.

\bibliographystyle{elsarticle-harv}
\bibliography{references}

@book{box2015time,
  author    = {Box, George E. P. and Jenkins, Gwilym M. and Reinsel, Gwilym C. and Ljung, Greta M.},
  title     = {Time Series Analysis: Forecasting and Control},
  edition   = {5},
  publisher = {John Wiley \& Sons},
  address   = {Hoboken, NJ},
  year      = {2015}
}

@article{cliver2022extreme,
  author  = {Cliver, Edward W. and Schrijver, Carolus J. and Shibata, Kazunari and Usoskin, Ilya G.},
  title   = {Extreme solar events},
  journal = {Living Reviews in Solar Physics},
  volume  = {19},
  number  = {2},
  year    = {2022},
  doi     = {10.1007/s41116-022-00033-8}
}

@misc{dataset,
  author    = {Woods, Thomas N.},
  title     = {{SORCE XPS Level 3 Solar Spectral Irradiance Daily Means V012}},
  publisher = {NASA Goddard Earth Sciences Data and Information Services Center (GES DISC)},
  address   = {Greenbelt, MD, USA},
  year      = {2020},
  doi       = {10.5067/ZBKF34FGDCLT},
  note      = {Digital science data}
}

@book{embrechts1997modelling,
  author    = {Embrechts, Paul and Kl{\"u}ppelberg, Claudia and Mikosch, Thomas},
  title     = {Modelling Extremal Events: For Insurance and Finance},
  publisher = {Springer},
  address   = {Berlin, Heidelberg},
  year      = {1997},
  doi       = {10.1007/978-3-642-33483-2}
}

@article{fisher1928limiting,
  author  = {Fisher, Ronald A. and Tippett, Leonard H. C.},
  title   = {Limiting forms of the frequency distribution of the largest or smallest member of a sample},
  journal = {Mathematical Proceedings of the Cambridge Philosophical Society},
  volume  = {24},
  number  = {2},
  pages   = {180--190},
  year    = {1928},
  doi     = {10.1017/S0305004100015681}
}

@article{frohlich2006solar,
  author  = {Fr{\"o}hlich, Claus},
  title   = {Solar irradiance variability since 1978},
  journal = {Space Science Reviews},
  volume  = {125},
  pages   = {53--65},
  year    = {2006},
  doi     = {10.1007/s11214-006-9046-5}
}

@article{gilleland2016extremes,
  author  = {Gilleland, Eric and Katz, Richard W.},
  title   = {{extRemes} 2.0: An extreme value analysis package in {R}},
  journal = {Journal of Statistical Software},
  volume  = {72},
  number  = {8},
  pages   = {1--39},
  year    = {2016},
  doi     = {10.18637/jss.v072.i08}
}

@article{gnedenko1943,
  author  = {Gnedenko, Boris V.},
  title   = {Sur la distribution limite du terme maximum d'une s{\'e}rie al{\'e}atoire},
  journal = {Annals of Mathematics},
  volume  = {44},
  number  = {3},
  pages   = {423--453},
  year    = {1943},
  doi     = {10.2307/1968974}
}

@article{gomez2018irradiancia,
  author  = {Rodr{\'i}guez G{\'o}mez, Jos{\'e} Mar{\'i}a and Carlesso, Franciele and Vieira, Luis Eduardo Antunes and da Silva, Luiz Antonio},
  title   = {A irradi{\^a}ncia solar: conceitos b{\'a}sicos},
  journal = {Revista Brasileira de Ensino de F{\'i}sica},
  volume  = {40},
  number  = {3},
  pages   = {e3312},
  year    = {2018}
}

@article{jenkinson1955frequency,
  author  = {Jenkinson, Arthur F.},
  title   = {The frequency distribution of the annual maximum or minimum values of meteorological elements},
  journal = {Quarterly Journal of the Royal Meteorological Society},
  volume  = {81},
  number  = {348},
  pages   = {158--171},
  year    = {1955},
  doi     = {10.1002/qj.49708134804}
}

@book{kotz2000extreme,
  author    = {Kotz, Samuel and Nadarajah, Saralees},
  title     = {Extreme Value Distributions: Theory and Applications},
  publisher = {World Scientific},
  address   = {Singapore},
  year      = {2000}
}

@article{ljung1978,
  author  = {Ljung, Greta M. and Box, George E. P.},
  title   = {On a measure of lack of fit in time series models},
  journal = {Biometrika},
  volume  = {65},
  number  = {2},
  pages   = {297--303},
  year    = {1978},
  doi     = {10.1093/biomet/65.2.297}
}

@article{rottman2005sorce,
  author  = {Rottman, Gary},
  title   = {The {SORCE} mission},
  journal = {Solar Physics},
  volume  = {230},
  pages   = {7--25},
  year    = {2005},
  doi     = {10.1007/s11207-005-8112-6}
}

@article{tsurutani2006extreme,
  author  = {Tsurutani, Bruce T. and Mannucci, Anthony J. and Iijima, Byron A. and Guarnieri, Fernando L. and Gonzalez, Walter D. and Judge, Darrell L. and Gangopadhyay, Paramita and Pap, Judit},
  title   = {The extreme Halloween 2003 solar flares and resultant extreme ionospheric effects: Comparison to other Halloween events and the Bastille Day event},
  journal = {Advances in Space Research},
  volume  = {37},
  number  = {9},
  pages   = {1583--1588},
  year    = {2006},
  doi     = {10.1016/j.asr.2005.05.114}
}

@article{von1936distribution,
  author  = {{von Mises}, Richard},
  title   = {La distribution de la plus grande de $n$ valeurs},
  journal = {Revue Math{\'e}matique de l'Union Interbalkanique},
  volume  = {1},
  pages   = {141--160},
  year    = {1936}
}

@article{watari2001bastille,
  author  = {Watari, Shinichi and Kunitake, Manabu and Watanabe, Takashi},
  title   = {The Bastille Day 14 July 2000 event in historical large Sun--Earth connection events},
  journal = {Solar Physics},
  volume  = {204},
  pages   = {425--438},
  year    = {2001},
  doi     = {10.1023/A:1014273227639}
}

@manual{wuertz2009package,
  author = {Wuertz, Diethelm and Setz, Tobias and Chalabi, Yohan},
  title  = {{fExtremes}: Rmetrics -- Modelling Extreme Events in Finance},
  year   = {2009},
  note   = {R package}
}

@article{woods2021overview,
  author  = {Woods, Thomas N. and Harder, Jerald W. and Kopp, Greg and McCabe, Debra and Rottman, Gary and Ryan, Sean and Snow, Martin},
  title   = {Overview of the Solar Radiation and Climate Experiment ({SORCE}) Seventeen-Year Mission},
  journal = {Solar Physics},
  volume  = {296},
  number  = {127},
  year    = {2021},
  doi     = {10.1007/s11207-021-01869-3}
}

@article{woods2005xps_overview,
  author  = {Woods, Thomas N. and Rottman, Gary and Vest, Robert},
  title   = {{XUV} Photometer System ({XPS}): overview and calibrations},
  journal = {Solar Physics},
  volume  = {230},
  pages   = {345--374},
  year    = {2005},
  doi     = {10.1007/s11207-005-4119-2}
}

@article{woods2005xps_variations,
  author  = {Woods, Thomas N. and Rottman, Gary},
  title   = {{XUV} Photometer System ({XPS}): solar variations during the {SORCE} mission},
  journal = {Solar Physics},
  volume  = {230},
  pages   = {375--387},
  year    = {2005},
  doi     = {10.1007/s11207-005-2555-7}
}

@article{woods2022xpsfinal,
  author  = {Woods, Thomas N. and Elliott, James},
  title   = {Solar Radiation and Climate Experiment ({SORCE}) {X}-Ray Photometer System ({XPS}): final data-processing algorithms},
  journal = {Solar Physics},
  volume  = {297},
  number  = {64},
  year    = {2022},
  doi     = {10.1007/s11207-022-01997-4}
}

@article{greenwood1979probability,
  author  = {Greenwood, J. Arthur and Landwehr, J. Maciunas and Matalas, N. C. and Wallis, J. R.},
  title   = {Probability weighted moments: Definition and relation to parameters of several distributions expressable in inverse form},
  journal = {Water Resources Research},
  volume  = {15},
  number  = {5},
  pages   = {1049--1054},
  year    = {1979},
  doi     = {10.1029/WR015i005p01049}
}

@article{hosking1985estimation,
  author  = {Hosking, J. R. M. and Wallis, J. R. and Wood, E. F.},
  title   = {Estimation of the Generalized Extreme-Value Distribution by the Method of Probability-Weighted Moments},
  journal = {Technometrics},
  volume  = {27},
  number  = {3},
  pages   = {251--261},
  year    = {1985},
  doi     = {10.1080/00401706.1985.10488049}
}

@article{del2018solar,
  title={Solar UV and X-ray spectral diagnostics},
  author={Del Zanna, Giulio and Mason, Helen E},
  journal={Living Reviews in Solar Physics},
  volume={15},
  number={1},
  pages={5},
  year={2018},
  publisher={Springer}
}

@article{phillips1945ultraviolet,
  title={Ultraviolet and X-ray Spectroscopy of the Solar Atmosphere},
  author={Phillips, Kenneth JH and Feldman, Uri and Landi, Enrico},
  journal={American history},
  volume={1861},
  number={1900},
  year={1945}
}

@book{aschwanden2006physics,
  title={Physics of the solar corona: an introduction with problems and solutions},
  author={Aschwanden, Markus J},
  year={2006},
  publisher={Springer}
}

@manual{R_base,
  title        = {R: A Language and Environment for Statistical Computing},
  author       = {{R Core Team}},
  organization = {R Foundation for Statistical Computing},
  address      = {Vienna, Austria},
  year         = 2025,
  adsurl       = {https://www.R-project.org/},
    url = {https://www.R-project.org/},
  note         = {Version 4.5.1}
}

@article{lepot2017interpolation,
  title={Interpolation in time series: An introductive overview of existing methods, their performance criteria and uncertainty assessment},
  author={Lepot, Mathieu and Aubin, Jean-Baptiste and Clemens, Fran{\c{c}}ois HLR},
  journal={Water},
  volume={9},
  number={10},
  pages={796},
  year={2017},
  publisher={MDPI},
  adsurl = {https://www.mdpi.com/2073-4441/9/10/796}
}

@book{hamilton2020time,
  title={Time series analysis},
  author={Hamilton, James D},
  year={2020},
  publisher={Princeton university press},
adsurl = {https://books.google.com.br/books?hl=pt-BR&lr=&id=BeryDwAAQBAJ&oi=fnd&pg=PP1&dq=%22Time+Series+Analysis%22+by+James+D.+Hamilton&ots=BhzS424Rip&sig=-Vx46NwHKv81-51ppzA5X56ijYU}
}

@techreport{jcgm2008gum,
  author      = {{Joint Committee for Guides in Metrology}},
  title       = {Evaluation of Measurement Data: Guide to the Expression of Uncertainty in Measurement},
  institution = {JCGM},
  year        = {2008},
  number      = {JCGM 100:2008}
}

@book{fuller1987measurement,
  author    = {Fuller, Wayne A.},
  title     = {Measurement Error Models},
  publisher = {John Wiley \& Sons},
  year      = {1987},
  doi       = {10.1002/9780470316665}
}

@book{foukal2008solar,
  title={Solar astrophysics},
  author={Foukal, Peter V},
  year={2008},
  publisher={John Wiley \& Sons}
}

@article{eadie2019realizing,
  title={Realizing the potential of astrostatistics and astroinformatics},
  author={Eadie, Gwendolyn and Loredo, Thomas J and Mahabal, Ashish A and Siemiginowska, Aneta and Feigelson, Eric and Ford, Eric B and Djorgovski, SG and Graham, Matthew and Ivezic, Zeljko and Borne, Kirk and others},
  journal={arXiv preprint arXiv:1909.11714},
  year={2019}
}

@book{decarvalho2026handbook,
  editor    = {de Carvalho, Miguel and Huser, Rapha{\"e}l
               and Naveau, Philippe and Reich, Brian J.},
  title     = {Handbook of Statistics of Extremes},
  year      = {2026},
  edition   = {1},
  publisher = {Chapman and Hall/CRC},
  address   = {Boca Raton, FL},
  series    = {Chapman \& Hall/CRC Handbooks of Modern Statistical Methods},
  isbn      = {978-1-032-51980-7},
  doi       = {10.1201/9781003404743},
  note      = {Forthcoming}
}

\end{document}